# Derivation of statistics of real delay time from statistics of imaginary delay time using spectroscopic technique in weakly disordered optical media

Prabhakar Pradhan,[1]* Peeyush Sahay,[2] and Huda M. Almabadi[2]

*Department of Physics and Astronomy, Mississippi State University,
Mississippi State, MS 39762*
*Department of Physics, University of Memphis, TN 38152*

*E-mail:PPradhan: pp838@msstate.edu

Abstract

Delay time is defined as a time that a wave spent in a scattering medium before it escapes, and this can be derived by the energy derivative of the phase of the scattering wave.. Considering a complex reflection coefficient $R = |R| exp(i\theta)$ of a light wave (wave vector: $k$), a study of real delay time ($\tau_r$) defined as, $\tau_r = d\theta/cdk$, and an 'imaginary delay time', defined as, $\tau_i = d\theta_i/cdk$, where $|R| = exp(-\theta_i))$, shows a strikingly similar statistical distribution, but with a shift in time, in a weak disorder and short sample length regime. Consequently, the statistics of $\tau_r$ can be derived from the reflection intensity of the wave in this regime. The range of the scattering parameters for this validation, as well as application of this method in characterization of weakly disordered media is also discussed.

## 1. Introduction:

*1.1 Delay time* Probing the optical properties of a sample via its light backscattering properties has been a major topics of research for a longtime. The complex reflection coefficient, $R = r^{1/2} exp(i\theta)$ has two parts: the absolute value $r^{1/2}$ and the associated phase $\theta$. In general, the reflection amplitude ($r^{1/2}$), or the intensity $RR^*$ (= $r$), has no phase information. In order to obtain the phase information, one requires complex reflection amplitude form to be measured [1]. In real case scenario or in an experimental situation, in most cases interference experiments must be performed simultaneously with respect to the phase of a reference signal to keep track of the phase of the reflected waves [2]. For example, systems such as Michelson's interferometry, optical coherence tomography (OCT), etc., based applications require a separate reference signal arm to monitor the changes in the phase [1,3]. In this context, we asked, can the phase information be extracted from the reflection intensity (*r*) itself, in the case when apparent phase information is missing? If such a case were to present itself, we further asked how to obtain information of any other physical parameters that depend on the phase or its derivatives? In particular, we focused on the Wigner delay time. Wigner delay time is defined as the time spent by an electron in an electronic scattering medium before it escapes from it [4]. In particular, it is defined as the energy derivative of the phase of the reflection coefficient, $\hbar d\theta/dE$. As $E=\hbar\omega$ for electronic case, the delay time also can be written as the derivative of the phase with respect to the angular frequency (*ω*), that is $d\theta/d\omega$. The definition of the delay time has been further generalized to the classical wave, such as electromagnetic waves, in the optical media. In this case, the delay time is defined as $d\theta/d\omega$ or $d\theta/cdk$, where *c* represents the speed of light [5]. Therefore, for classical waves, the Wigner delay time can be considered as the spectral derivative (as $k=2\pi/\lambda$) of the phase. Using this definition of the delay time, in this paper we will derive the connection of real and 'imaginary delay time' defined later in the section.

Properties of Wigner delay time have been extensively studied for both electronics and optics for larger sample length [4-18]. While a majority of the work have focused on studying the delay time distributions in a random passive medium (no absorption or amplification) at different length scales, some of the work have also focused on studying the delay times in active medium (absorption/ amplification allowed), by considering systems ranging from one dimensional disordered media to chaotic cavities. One of the important parameter which is often defined for the sample in such studies is the scattering strength of the sample *L/ξ*, where *L* is the sample length and *ξ* is its localization length. Studies suggest that from *L/ξ << 1* (i.e., Ballistic regime) to *L/ξ >> 1* (i.e., localized regime) the delay time statistics shifts from Gaussian to log-normal especially for the larger delay times in 1D disordered media [12]. It should be noted that, in the localized regime (*L/ξ >> 1*), the analytical expression of the probability distribution is known [12].

In this work, we have considered a weak disorder and short sample length ($L/\xi << 1$) system, and have studied the real delay time (i.e., the Wigner delay time, which is derived from the real phase of the scattered wave) and 'imaginary delay time' which is derived from an 'imaginary phase' obtained from the amplitude part of the scattered wave. The results show that both have similar statistical form, but with a shift in time. As pointed above, the phase information is generally lost in the intensity expression (because the intensity is the square of the absolute value of the complex reflection amplitude), however, in this work, we show that the real delay time statistics for a weakly disordered media with $L << \xi$, can be recovered from the reflected intensity statistics, without actually performing phase measurements. In essence, we show that the imaginary delay time acts as the counter for the real delay time and can be used to extract the phase information for such systems.

Because the Wigner delay time ($\tau_r$) is calculated from the real phase of the reflected wave [5], consider a complex reflection coefficient $R$ of the light waves and $\tau_r$ represented as follows:

$$R = r^{1/2} e^{i\theta}, \qquad (1)$$

$$\tau_r = \frac{d\theta}{d\omega} = \frac{d\theta(k)}{cdk} \equiv \frac{d\theta(k)}{dk}. \qquad (2)$$

In the above equations, $\omega$ is the angular frequency, $k$ is the wavenumber, and $c$ represents the speed of light in vacuum. To simplify the calculation, we will consider $c = 1$. Note that, in disordered optical medium, $\tau_r$ is a statistical quantity similar to the reflection coefficient. It should be highlighted that, in this paper we have consistently referred $\tau_r$ as 'real delay time', for the reason that this quantity is derived from the real phase of the scattered waves, and importantly to differentiate it from the 'imaginary delay time' which has been represented as $\tau_i$. We define an 'imaginary delay time', which is calculated from the magnitude of the reflection amplitude $|R|$ by introducing an 'imaginary phase' ($\theta_i$) defined as:

$$|R| = r^{1/2} = e^{-\theta_i} = e^{i(i\theta_i)} \qquad (3)$$

$$\theta_i = -\ln(|R|) \qquad (4)$$

$$\tau_i \equiv \frac{d\theta_i(k)}{d\omega} \propto \frac{d\theta_i(k)}{d(ck)} = -\frac{1}{|R|} \frac{d|R|}{dk} \qquad (5)$$

Consequently, an analogy between the real and imaginary delay time can then be shown as follows. For a complex reflection coefficient $R$,

$$R = |R|e^{i\theta} = e^{-\theta_i}e^{i\theta} = e^{i(i\theta_i)}e^{i\theta},\tag{6}$$

$$\tau_r = \frac{d\theta}{dk} \text{ and } \tau_i = \frac{d\theta_i}{dk}.\tag{7}$$

*1.2. Derivation of real and imaginary delay times: $\tau_r$ and $\tau_i$, from the Langevin equations of r and $\theta$.* We numerically calculated the real delay time distribution $\tau_r$, and the imaginary delay time $\tau_i$, and their statistical properties to examine any correlation between them. Consider a light wave of wave vector $k$ (i.e., $k = 2\pi/\lambda$, $\lambda$: wavelength of the light) reflected from a one-dimensional (1D) disordered optical medium of refractive index $n(x) = n_0 + dn(x)$ and length $L$. In this case, a Langevin equation for the complex reflection coefficient $R(L)$ can be derived as follows [19-21]:

$$\frac{dR(L)}{dL} = 2ikR(L) + \frac{ik}{2}\frac{2dn(L)}{n_0}[1+R(L)]^2,\tag{8}$$

with the initial condition $R(L=0) = 0$. Now consider,

$$R(L) = |R(L)|\exp(i\theta(L)) = r(L)^{1/2}\exp(i\theta(L)).\tag{9}$$

Using the above relation for $R(L)$, in Eq. (8), we obtain two coupled differential equations in $r$ and $\theta$ [19]:

$$\frac{dr(L)}{dL} = f_r(r,\theta;L) = k\frac{2dn}{n_0}r^{1/2}(1-r)\sin\theta,\tag{10}$$

$$\frac{d\theta(L)}{dL} = f_\theta(r,\theta;L) = 2k + \frac{k}{2}\frac{2dn}{n_0}[2+(r^{1/2}+r^{-1/2})\cos\theta].\tag{11}$$

Using Eqs. (10) and (11), $\tau_r$ and $\tau_i$ can also be expressed as, ($c = 1$):

$$\tau_r \equiv \frac{d\theta_r(\omega)}{d\omega} = \frac{d\theta_r(k)}{dk} = \frac{d}{dk}\int_0^L f_\theta(r(k),\theta(k);L')dL',\tag{12}$$

$$\tau_i \equiv \frac{d\theta_i(\omega)}{d\omega} = \frac{d\theta_i(k)}{dk} = \frac{1}{|R|}\frac{d}{dk}\int_0^L f_r(r(k),\theta(k);L')dL',\tag{13}$$

**2. Numerical simulations for P($\tau_i$) and P($\tau_r$) statistics:**

We performed numerical simulations for Eq. (9) to determine $R(L)$, and subsequently to calculate the values of $\tau_r$ and $\tau_i$. Simulations were performed for various sample lengths $L = 1 - 40\ \mu m$, over the visible light range, $\lambda = 0.5 - 0.7\ \mu m$. Keeping practical applications in mind, all studies were performed by considering the refractive index fluctuation strength $dn = 0.02$, which is relevant to biological systems [22-24]. Furthermore, to better mimic a natural biological sample, we considered the Gaussian color noise model of refractive index fluctuations with short range exponential decay spatial correlation of correlation length $l_c \sim 0.02\ \mu m$ [23,24]. The localization length for such a system can be calculated as, $\xi = (2dn^2 k^2 l_c)^{-1} = 570\ \mu m$ at $\lambda = 0.6\ \mu m$. Correspondingly, such a system represents a weak disordered optical system with $L/\xi \leq 0.07$, i.e., short sample length scale. It should be noted that for such a system, the term $L/\xi$ also represents the average reflection coefficient ($r_a$) of the system, i.e., $r_a = L/\xi$ [19,24].

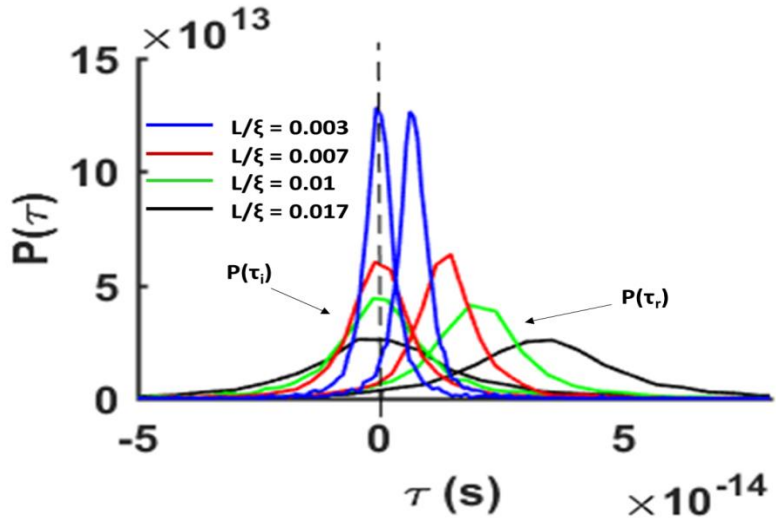

FIG.1. $P(\tau_r)$ and $P(\tau_i)$ distributions for reflection coefficient $r_a = L/\xi = 0.003 - 0.017$ by the direct simulation of the stochastic Langevin equations.

***2.1. $P(\tau_i)$ and $P(\tau_r)$ statistics:*** Figure 1 shows plots of the normalized probability distribution of $\tau_r$ and $\tau_i$, for four different $L/\xi$ values. For each $L/\xi$, it can be seen that the distribution of $P(\tau_r)$ (right) has similar shape as that of $P(\tau_i)$ (left), but with a shift in time. Interestingly, both $P(\tau_r)$ and $P(\tau_i)$ appears to have similar standard deviation (*std*) as well, with their distribution symmetric around their respective mean. It can also be noted that while the mean of $P(\tau_i)$ remains at zero, the mean of $P(\tau_r)$ increases with the increase in the $L/\xi$ values.

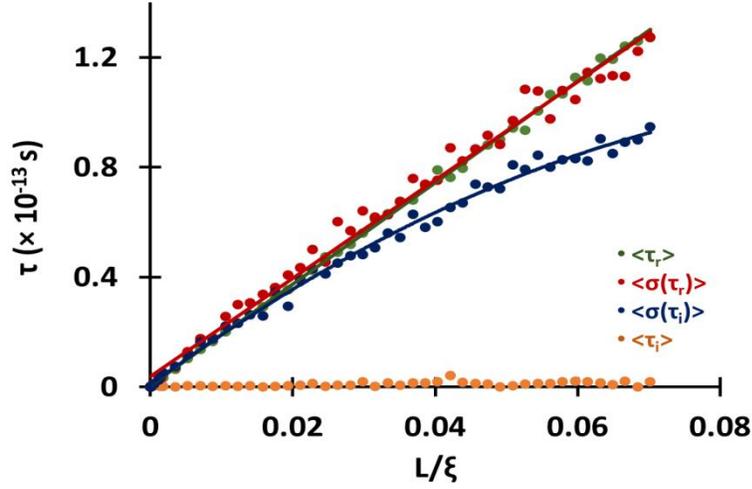

FIG. 2. The variation of the mean and *std* of the real and imaginary delay time as a function of $L/\xi$ (= 0 – 0.07). The $<\tau_r(L)>$, $\sigma(\tau_r(L))$ and $\sigma(\tau_i(L))$, follows approximately same variation as a function of the $L/\xi$ or the reflection intensity till $L/\xi \sim .03$. However $\sigma(\tau_i(L))$ start deviating after this point, but the value of $<\tau_i(L)>$ is always zero.

The Fig. 2 shows the simulation results, carried out at the above mentioned parameters, by plotting the variation of $<\tau_r(L)>$, $<\tau_i(L)>$, $\sigma(\tau_r(L))$, and $\sigma(\tau_i(L))$ at different values of $L/\xi = 0 -$ 0.07. The results show that the $<\tau_r(L)>$, $\sigma(\tau_r(L))$, and $\sigma(\tau_i(L))$ values increase with increase in $L/\xi$ value. While up to $L/\xi \sim 0.03$, the linearity of these three parameters, ( i.e., $<\tau_r(L)>$, $\sigma(\tau_r(L))$, and $\sigma(\tau_i(L))$), remains same, after this point the $\sigma(\tau_i(L))$ starts deviating with increasing $L/\xi$ value. The $<\tau_i(L)>$ values, however, remains zero for all the $L/\xi$ values. This result is significant. The same linear dependence of at least three of the parameters, namely $<\tau_r(L)>$, $\sigma(\tau_r(L))$, and $\sigma(\tau_i(L))$, till $L/\xi \sim 0.03$, suggests one parameter characterization theory of the delay time– i.e., these quantities can be characterized by only one parameters, in this length scale range. Therefore, it further also suggests that this relationship can be exploited to reconstruct real delay time statistics from the imaginary delay time statistics, which is obtained from the intensity measurement in an actual experiment.

**3. Deriving real delay time information from the imaginary delay time:**

By understanding the statistics of imaginary delay time distribution, the real delay time statistics can be constructed. In particular, as shown in Figs. 1 and 2, as well as described in the Sec. 2.1, we can see that $P(\tau_r)$ and $P(\tau_i)$ are related by a shift in time of amount $\sigma(\tau_i)$, that is:

$$P(\tau_r) \approx P(\tau_i + \sigma(\tau_i)).  \qquad (14)$$

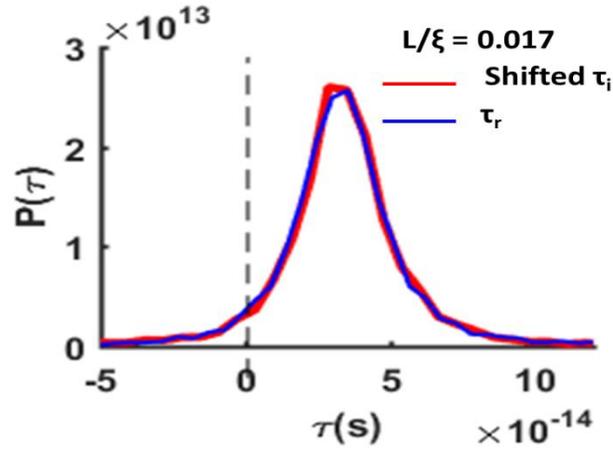

FIG. 3. $P(\tau_r)$ derived from $P(\tau_i)$, by shifting $P(\tau_i)$ by its std value $\sigma(\tau_i)$; plotted at $L/\xi$ (= 0.017). The plots are are well overlapped on each other. All plots $L/\xi$ <0.03 will behave the same nature.

Figure 3 demonstrate the relation shown in Eq. (14) for one such representative example, by plotting $P(\tau_r)$ and $P(\tau_i + \sigma(\tau_i(L)))$ at $L/\xi \sim 0.017$, corresponding to without shift in Fig.1 for the same parameter. We plotted the probability distribution of $\tau_r$ and $\tau_i$, with $\tau_i$ values shifted by its standard deviation value. As can be seen in Fig. 3, which shows a very well overlapping between the real and the imaginary delay time statistics obtained in such a way.

This relationship is valid for a range of the parameter space in $L$ and $\xi$, for the average reflection coefficient $r_a(L,\xi)$. Our work mainly considers weak disorder regime- one that is particularly relevant to biological systems, but it can be generalized for any optical disordered sample, as long as $r_a < 0.03$. In this case, the valid range of the weak disorder is around $kL \sim (2\pi/.6) \times 20 \sim 210$ and $L/\xi \sim 0.03$. This is a wide sample reflection range for mesoscopic disordered optical samples and many optical biological samples, such as cells and tissues exists.

## 4. Asymptotic limit verification:

To verify the results of our numerical simulation, we computed delay time distribution at asymptotic limit (larger sample lengths), where the analytical expression for $P(\tau_r)$ is readily known. In this limit, it was shown that $\tau_r$ has a functional form, as presented in Eq. (15) [11-12].

$$P(\tau_r) = \frac{\alpha}{\tau_r^2} \exp\left(-\frac{\alpha}{\tau_r}\right) \tag{15}$$

Where $\alpha = 4(dn^2 k)^{-1}$ and $dn^2$ and $k$ parameters are defined above. We performed numerical simulations of $P(\tau_r)$ with the asymptotic limit defined as large length ($L=20\ \mu m$) and variation of the refractive index $dn = 0.2$ with short correlation length $l_c = 0.02\ \mu m$. The numerical simulation results with the analytical results are shown in Fig. 4, where we have defined $\tau_s = \alpha/\tau_r$. As it can be seen in the plot that in this asymptotic limit, both numerical simulation result and analytical form match well, validating our numerical delay time calculation algorithm used here for the work.

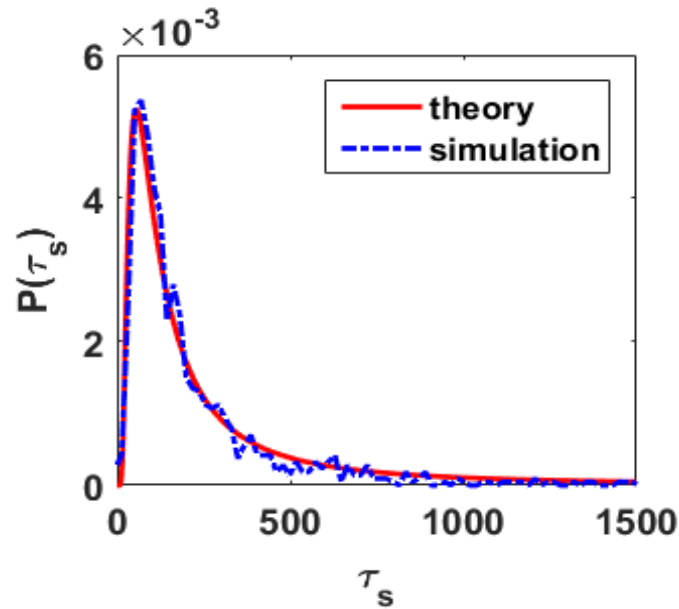

FIG. 4. Verification of our numerical simulation analytical results for large length limit. With $\tau_s$ as a scaled quantity, $\tau_s = \alpha/\tau_r$, plots of $P(\tau_s)$ calculated by simulation and theory in asymptotic limit ($L=20\ \mu m$, $dn=0.2$)

## 5. Conclusion and Discussion:

In this paper, we numerically studied the statistics of the delay times for one-dimensional weakly disordered optical media with short sample lengths. The statistics of real delay time ($\tau_r$) i.e., Wigner delay time, which is derived from the spectroscopic derivative of the phase $\theta$ of a reflection wave $R = r^{1/2} exp(i\theta)$, was compared with the statistics of an 'imaginary delay time' ($\tau_i$) derived from the spectroscopic derivative of the reflection amplitude $|R|$, in visible range. The results show a striking similarity between their probability distributions, i.e., $P(\tau_r)$ and $P(\tau_i)$. In particular, one distribution can be projected on other by an appropriate shift of standard deviation values. The average real delay time $<\tau_r(L)>$, and the standard deviation of the real and imaginary delay times (i.e., $\sigma(\tau_r(L))$, and $\sigma(\tau_i(L))$ respectively) increases linearly with $L/\xi$, with a same slope up to $L/\xi \sim 0.03$, after which the $\sigma(\tau_i(L))$ values starts deviating. The average imaginary delay time $<\tau_i(L)>$ remains zero for the all the $L/\xi$ values. These results suggested that we can extract the statistical properties of the real delay time distribution (which in actual practice requires phase measurement) from the imaginary delay time distribution, obtained from intensity information, at least for a range of $L/\xi$ values ($L/\xi < 0.03$). It should be noted that $L/\xi < 0.03$ is a wide range of parameter space especially for a weakly disordered regime and short sample lengths, for back scattering from optical mesoscopic samples. It is worth mentioning here that while the $<\tau_r(L)>$ and $\sigma(\tau_r(L))$ values are expected to increase with increase in the sample length ($L$) and the disorder strength ($dn$) and therefore with increase in $L/\xi$ values. The respective patterns of $<\tau_i(L)>$ and $\sigma(\tau_i(L))$ observed in the present study can be attributed to the fluctuating nature of the reflection coefficient ($r$) with respect to wavenumber ($k$) of the light wave incident on such a disordered system. The results for $P(\tau_r)$ are consistent with approximate analytical results [13]. As the $L/\xi$ value increases, all the above statistical averages saturates eventually (not shown here), the reason for which can be attributed to the saturation in the reflection intensity at a larger $L/\xi$ values, which reaches maximum to 1.

Many examples of weak disorder systems can be cited, varying from biological systems to polymers. For instance, biological samples such as cells and tissues are typical examples of weakly disordered media. As we have shown, in this regime of weak disorder and short length, reflection amplitude information is sufficient to derive the real delay time distribution, without actually deriving the real phase and its derivative. Therefore, this technique potentially may add a new dimension to the understanding of phase statistics, as well as delay time, for weakly disordered media and samples of short lengths. From application point of view, at least two potential applications are discussed here: 1) Phase information from intensity information; and 2) Biological sample characterization from their light reflection statistics and its applications to disease detection. There are lots of cases where we require delay time statistics data, however the phase statistics is not easy to obtain as it requires detailed interference experiments. Therefore, this method would be very much useful for deriving the delay time statistics from the reflection intensity spectra. It was recently shown that the intracellular structural disorder increase in the cell in progressive carcinogenesis, due to the nanoscale alteration of the basic building blocks of the cells, such as DNA, protein, and lipids [23-24]. Since the delay time (i.e., Wigner delay time) provides a measure of extra time taken by photons to scatter from a media, it may be possible to use the delay time approach to differentiate normal and cancerous cell from their backscattering spectra, as in the case of progressive carcinogenesis the disorder level in the cells is known to increase, due to the nano-alterations as described above. Thus, this increased disorder would lead to higher delay time for these samples. As the method consider the spectral derivation of the phase, therefore, the delay time as a parameter is more sensitive relative to the direct phase changes in the reflection intensity spectra. Conversely, the imaginary delay time which is easy to determine can be used to characterize the disordered cancer cells as a counter to the real delay time. Cancer cells are an example of the natural weakly disordered system where the nano-to submicron scale structural disorder increases with the progress of carcinogenesis. In fact it may be possible to utilize imaginary delay time statistics as a new marker, or parameter, to characterize structural properties of biological cells or similar weakly disordered natural and artificially made samples..

*Acknowledgements:* The work was supported by the NIH (R01EB003682, R01EB016983) and Mississippi State University. We thank D. J. Park for useful discussions.